\documentclass[journal,twoside]{IEEEtran}

\usepackage[T1]{fontenc}
\usepackage[utf8]{inputenc}
\usepackage{cite}
\usepackage{amsmath,amssymb}
\usepackage{graphicx}
\usepackage{url}
\usepackage{booktabs}
\usepackage{array}
\usepackage{tabularx}
\usepackage{multirow}
\usepackage{balance}
\usepackage{microtype}
\usepackage{xcolor}

\newcolumntype{P}[1]{>{\raggedright\arraybackslash}p{#1}}

\begin{document}

\title{Sovereign 2.0: Control-Plane Sovereignty for Cloud Systems Under Disruption}

\author{
\begin{tabular}{cc}
Justin Stark & Scott Wilkie \\
University of Technology Sydney & University of Technology Sydney \\
Sydney, NSW, Australia & Sydney, NSW, Australia \\
Accenture, Australia & Accenture, Australia \\
justin.stark@uts.edu.au & scott.wilkie@uts.edu.au \\
justin.stark@accenture.com & scott.wilkie@accenture.com
\end{tabular}
}

\maketitle

\begin{abstract}
Cloud sovereignty can no longer be defined by data residency or infrastructure location alone. Under conditions of geopolitical disruption, legal exposure, and expanding service boundaries, sovereignty must be understood as enforceable control over how digital services are governed, operated, and recovered.

This paper introduces \textit{Sovereign 2.0}, a control-plane-centric model that extends sovereignty beyond localisation to include governance authority, privileged access, cryptographic trust, data lifecycle control, observability, and incident response across federated environments. We define \textit{management sovereignty} as the sovereign ability to govern, operate, evidence, and recover services regardless of underlying infrastructure dependencies.

To operationalise this model, we propose a three-layer risk-assurance framework spanning governance, operational, and technical controls, enabling sovereign outcomes to be specified and continuously evidenced under both steady-state and crisis conditions. We further position post-quantum-ready cryptographic control, particularly TLS and key custody, as foundational to long-term sovereign trust.

These contributions reframe sovereignty as an evidence-backed control system rather than a property of location, with implications for cloud architecture, procurement, and resilience design.
\end{abstract}

\begin{IEEEkeywords}
cloud sovereignty, digital sovereignty, sovereign cloud, risk assurance, operational resilience, data residency, cloud guardrails, post-quantum cryptography, TLS~1.3, hybrid key exchange
\end{IEEEkeywords}


\section{Introduction}

\IEEEPARstart{C}{\lowercase{loud}} sovereignty has traditionally been framed as a question of location: where data is stored, where infrastructure resides, and which jurisdiction governs the provider. This framing—here termed \textit{Sovereign 1.0}—has shaped regulatory policy, procurement requirements, and cloud architecture design for over a decade. However, it is no longer sufficient to describe or assure sovereign control in modern cloud systems.

Three developments expose a structural gap between residency-based sovereignty and contemporary service delivery. First, geopolitical and kinetic disruptions demonstrate that cloud infrastructure remains physically vulnerable, and that regional concentration can produce correlated service degradation. Second, extraterritorial legal regimes challenge the assumption that data location alone determines control. Third, the effective service boundary has expanded beyond the data centre to include identity systems, observability pipelines, SaaS dependencies, AI services, and remote support pathways. As a result, systems can satisfy residency requirements while still lacking sovereign control over how they are governed, operated, and recovered.

These conditions invalidate a purely location-centric model of sovereignty. The relevant question is no longer where data resides, but whether control over the service—including governance authority, privileged access, cryptographic trust, data lifecycle, observability, and incident response—remains enforceable within a sovereign boundary under both steady-state and disruption conditions.

This paper defines \textit{Sovereign 2.0}, a control-plane-centric model that reframes sovereignty as an enforceable property of system control rather than infrastructure location. In this model, sovereignty is evaluated in terms of control retention: the extent to which governance authority, operational capability, evidence, and trust remain within the sovereign boundary under conditions of disruption.

We introduce \textit{management sovereignty} as the core operating construct of Sovereign 2.0: the sovereign ability to govern, approve, operate, evidence, and recover digital services across their full service boundary, even when underlying infrastructure and dependencies remain federated. This shifts sovereignty from a static compliance property to a dynamic, evidence-backed control system.

To operationalise this model, we propose a three-layer risk-assurance framework spanning governance, operational, and technical domains. This framework provides a structured mechanism to specify, implement, and continuously evidence sovereign outcomes, bridging the gap between policy intent and enforceable control in both steady-state and crisis conditions.

The significance of this shift is both practical and systemic. As cloud systems become more distributed, interdependent, and long-lived, sovereignty must be engineered as a control problem rather than a location property. This paper establishes that reframing and defines the foundation for an emerging discipline of sovereign service delivery.

\subsection{Contributions}

This paper makes four contributions.

\begin{itemize}
    \item We redefine cloud sovereignty for contemporary distributed systems by showing that residency- and localisation-based models are insufficient under conditions of geopolitical disruption, legal exposure, and service-boundary expansion.

    \item We introduce \textit{Sovereign 2.0}, a control-plane-centric model that reframes sovereignty as an enforceable property of service delivery rather than infrastructure location.

    \item We define \textit{management sovereignty} as the operating construct of Sovereign 2.0 and identify the control domains—governance, identity and privileged access, cryptographic trust, data lifecycle, observability, and incident authority—that must remain under sovereign control.

    \item We propose a three-layer risk-assurance framework spanning governance, operational, and technical controls, providing a structured mechanism to specify, implement, and continuously evidence sovereign outcomes across federated environments.
\end{itemize}

\section{Threat Model and Problem Context}

Sovereign service delivery must be evaluated under conditions where control over digital systems is contested, degraded, or disrupted. This paper considers a threat model that extends beyond conventional cybersecurity concerns to include geopolitical, legal, operational, and cryptographic stressors that affect the control plane of cloud-based services.

\subsection{System Model}

We consider a distributed service $S$ composed of infrastructure, platform, and application components deployed across one or more cloud environments. The effective service boundary of $S$ includes not only compute and storage resources, but also identity systems, observability pipelines, software delivery infrastructure, external SaaS dependencies, and provider-operated support pathways. As a result, $S$ is inherently federated, with control functions distributed across multiple administrative and jurisdictional domains.

\subsection{Adversarial and Disruption Model}

We define four classes of threats that directly impact sovereign control:

\textbf{1) Kinetic and Infrastructure Disruption.}
Cloud infrastructure is physically instantiated and therefore subject to damage, degradation, or denial of service due to conflict, natural disasters, or targeted attacks. Such events may cause partial or complete loss of regional capacity, requiring rapid migration or failover under constrained conditions.

\textbf{2) Jurisdictional and Legal Compulsion.}
Data and control planes may be subject to foreign legal regimes, including lawful access requests, disclosure obligations, or administrative control exerted through providers operating under different jurisdictions. These mechanisms can bypass or override local governance assumptions.

\textbf{3) Control-Plane Dependency and Supplier Failure.}
Critical control functions—such as identity, logging, key management, and administrative access—may depend on external providers or cross-border support arrangements. Loss, compromise, or unavailability of these dependencies can impair the ability to govern or recover the system.

\textbf{4) Cryptographic Degradation and Trust Erosion.}
Long-lived services rely on public-key cryptography for identity, confidentiality, and integrity. Emerging threats, including post-quantum adversaries and ``harvest now, decrypt later'' strategies, may undermine existing cryptographic mechanisms over time, weakening trust relationships across the service boundary.

\subsection{Sovereignty Failure Modes}

Under this threat model, a system fails to achieve sovereignty if any of the following conditions occur:

\begin{itemize}
    \item Loss of governance authority: the inability to define or enforce policy, approve actions, or exercise decision rights within the sovereign boundary.
    \item Loss of operational control: the inability to administer systems, execute recovery actions, or maintain service continuity without external or non-sovereign intervention.
    \item Loss of evidentiary control: the inability to access, verify, or retain logs, traces, and audit artefacts required to demonstrate compliance and accountability.
    \item Loss of trust control: the inability to govern cryptographic mechanisms, key material, or trust anchors that underpin system identity and secure communication.
\end{itemize}

These failure modes highlight that sovereignty is not solely a property of data location, but of sustained control over the mechanisms that govern, operate, and verify the service.

\subsection{Problem Statement}

Given a federated service $S$ operating under the threat conditions described above, the problem addressed in this paper is as follows:

\textit{How can sovereign authority over governance, operation, evidence, and trust be maintained and demonstrably enforced across the full service boundary of $S$, even when underlying infrastructure and dependencies are distributed across multiple jurisdictions and subject to disruption?}

This problem definition motivates the transition from location-centric sovereignty models to the control-plane-centric approach developed in the remainder of this paper.

\section{Why the Problem Is Now Time-Critical}
\label{sec:timecritical}

\subsection{Cloud Is Physical, Concentrated, and Therefore Disruptable}

The 2026 AWS Middle East incident converted an abstract policy concern into an operational one. AWS publicly reported direct and nearby drone-strike impacts on facilities in the UAE and Bahrain, significant impairment in two UAE Availability Zones, and degraded availability across foundational and dependent services~\cite{aws_rss_2026}. AP's reporting highlighted the same point from a different angle: cloud infrastructure remains a set of real physical facilities that can be damaged by conflict and cannot be made invisible simply because the service model is abstracted~\cite{ap_chan_2026}.

The implications are broader than a single outage. First, critical workloads are often concentrated into a small number of preferred regions for latency, ecosystem, and regulatory reasons. Second, resilience architectures are frequently optimised for failures within a region or a single Availability Zone, not for regional instability combined with evacuation, migration, and cross-border legal complexity. Third, recovery becomes a governance problem as much as a technical one: who has the authority to reroute traffic, relax policy constraints, activate alternate-region capacity, approve privileged changes, or invoke emergency supplier support?

\subsection{Residency Does Not Settle the Sovereignty Question}

A second reason for urgency is that residence and sovereignty are not the same. The Government of Canada has long distinguished data residency from data sovereignty, warning that data stored in cloud environments may still be subject to foreign laws regardless of where the infrastructure is physically located~\cite{gc_digital_sov_2025}. Canada's newer digital sovereignty framework extends the discussion further, defining digital sovereignty in terms of autonomy over digital assets and services, with explicit attention to operational resilience, system integrity, supply-chain factors, and institutional control~\cite{gc_digital_sov_2025}.

France's ANSSI makes a similar move through SecNumCloud. ANSSI states that SecNumCloud is intended to recognise trusted cloud offers suitable for sensitive data and notes that the qualification aims in particular to protect sensitive data and processing against cybercriminal threat and the application of extraterritorial laws~\cite{anssi_cloud_2026,anssi_secnumcloud_2026}. This is conceptually important: the target is not merely local hosting, but legally and operationally meaningful control.

\subsection{The Service Boundary Has Expanded Beyond the Data Centre}

The final reason the issue is time-critical is that sovereign exposure increasingly appears outside traditional hosting scopes. Telemetry pipelines may export logs and metadata to foreign-managed observability platforms. Identity federation may rely on external trust anchors. Managed support channels may create privileged administrative paths across jurisdictions. AI and analytics services may move sensitive prompts, embeddings, or intermediate data into external processing chains. Each of these can become a sovereignty leakage path unless it is explicitly governed.

For this reason, sovereign service delivery must be evaluated at the service boundary rather than only at the hosting boundary. The question is not just whether data starts inside the jurisdiction, but whether governance, identity, keys, evidence, and incident response remain under sovereign authority throughout the full operating lifecycle.

\section{From Sovereign 1.0 to Sovereign 2.0}
\label{sec:model}

A useful way to structure the sovereignty discussion is to begin with the now-common progression from sovereignty by localisation, to sovereignty by control, to sovereignty by design. This paper groups these into two eras.

\emph{Sovereign~1.0} is the location-centric model. It maps most closely to sovereignty by localisation and the traditional emphasis on in-country hosting, local facilities, and geographic data containment. It remains important, but on its own it mainly answers \emph{where} workloads sit.

\emph{Sovereign~2.0} is the management- and design-centric model. It incorporates sovereignty by control and sovereignty by design, shifting attention to \emph{who} can administer, access, approve, evidence, and recover services, and to whether architecture embeds sovereignty constraints such as strong segmentation, dedicated trust anchors, independent key control, and minimised dependence on foreign-operated control planes.

This distinction matters because real systems are inherently federated and layered. A platform can be localised at the data plane yet dependent on a foreign support path. It can be well controlled operationally yet lack sovereign recovery authority during crisis. It can be well designed cryptographically yet fail to provide locally governed evidence or supplier transparency. What is needed is a construct that focuses on the control outcomes that must remain sovereign across a mixed environment.

\subsection{Management Sovereignty}

This paper calls that construct \emph{management sovereignty}, defined as:

\begin{quote}
A system satisfies management sovereignty if governance, operational execution, evidence generation, and trust control remain enforceable within the sovereign boundary under both steady-state and disruption conditions.
\end{quote}

Management sovereignty is not the negation of Sovereign~1.0; rather, it operationalises Sovereign~2.0 by subsuming control and design questions into a single enforceable model. This emphasis resonates with recent policy language around ``Sovereignty~2.0,'' which rejects geographic abstractions in favour of measurable operational autonomy under selective interdependence~\cite{wef_2025}. A sovereign architecture need not eliminate all external dependencies; it must ensure that no single provider, jurisdiction, or support chain can unilaterally dictate outcomes during stress.

Under this definition, a service is more sovereign when the following remain under sovereign authority:
\begin{itemize}
  \item \emph{Governance and decision rights:} who can set policy, grant
        exceptions, and authorise emergency actions;
  \item \emph{Identity and privileged access:} who can administer systems,
        from where, under which approval chain, and with which evidence;
  \item \emph{Cryptographic trust:} who controls keys, certificate authorities,
        hardware security module (HSM) policies, and protocol transition
        decisions;
  \item \emph{Data lifecycle and egress:} how data is classified, tokenised,
        exported, replicated, and deleted;
  \item \emph{Observability and evidence:} where logs, traces, and alerts
        reside and who can inspect them; and
  \item \emph{Incident and continuity authority:} who commands recovery when
        the normal operating posture collapses.
\end{itemize}

This lens aligns with the Government of Canada's move from data sovereignty to digital sovereignty~\cite{gc_digital_sov_2025} while remaining compatible with more stringent national frameworks such as SecNumCloud~\cite{anssi_cloud_2026,anssi_secnumcloud_2026}. It also better reflects the operational lessons of the Middle East cloud disruption, where the hardest questions were not only about storage location, but about migration, control, recovery, and authority.

This control-plane perspective also extends to cryptographic trust. As long-lived services depend on public-key infrastructure and transport security, the ability to govern cryptographic transitions—particularly toward post-quantum-safe mechanisms—becomes part of sovereign control rather than a purely technical upgrade.

\section{Comparative Standards and Policy Landscape}
\label{sec:standards}

No single global sovereign-cloud standard exists. Instead, the policy landscape is evolving through a combination of certification schemes, security baselines, trusted-cloud qualifications, and operational control frameworks.
Table~\ref{tab:frameworks} summarises selected instruments relevant to
management sovereignty. The list is illustrative rather than exhaustive, but it shows a striking convergence around governance, legal exposure, identity, cryptography, auditability, and continuity.

\begin{table*}[!t]
\caption{Selected Sovereignty-Relevant Frameworks and Control Baselines
(Illustrative, Not Exhaustive)}
\label{tab:frameworks}
\centering
\renewcommand{\arraystretch}{1.25}
\begin{tabular}{P{2.0cm} P{2.6cm} P{3.5cm} P{7.4cm}}
\toprule
\textbf{Jurisdiction /}  &
\textbf{Instrument}      &
\textbf{Primary Emphasis}&
\textbf{Relevance to Management Sovereignty} \\
\textbf{Ecosystem} & & & \\
\midrule
European Union &
EU Cybersecurity Certification Framework; EUCS candidate
scheme~\cite{eu_ccf_2026,enisa_eucs_2020} &
EU-wide certification language, scheme governance, assurance levels
(basic, substantial, high) &
Creates a common assurance vocabulary across member states; useful for
procurement and comparability, but EUCS remains a candidate rather than a
finalised sovereign-cloud scheme. \\[4pt]

European Commission procurement &
Cloud Sovereignty Framework v1.2.1; EUR~180M sovereign-cloud
tender~\cite{ec_press_2025,ec_csf_2025} &
Eight sovereignty objectives; minimum SEAL thresholds; sovereignty scoring as
tender award input &
Turns sovereignty from declarative policy into measurable procurement and
assurance criteria, closely matching a Sovereign~2.0 and management-sovereignty
approach. \\[4pt]

Gaia-X ecosystem &
Gaia-X Trust Framework and
Label~\cite{gaiax_home_2026,gaiax_label_2024,gaiax_fw_2026} &
Transparency, portability, interoperability, compliance, ``European Control,''
federated trust &
Important for multi-party ecosystem governance where sovereignty depends on
verifiable policy and trust assertions across federated participants. \\[4pt]

France &
ANSSI SecNumCloud
v3.2~\cite{anssi_cloud_2026,anssi_secnumcloud_2026} &
High technical, operational, and legal requirements for trusted cloud;
sensitivity to extraterritorial-law exposure &
One of the clearest examples of sovereignty extending beyond hosting to legal
and operational conditions of trust. \\[4pt]

Germany &
BSI C5~\cite{bsi_c5_2026,bsi_c5_pdf_2020} &
Minimum cloud controls, independent audit, jurisdiction and disclosure
transparency, logging, key management, continuity &
Strong evidence-oriented baseline for enterprise cloud assurance; especially
useful for proving operational and technical control maturity. \\[4pt]

United Kingdom &
NCSC Cloud Security Principles~\cite{ncsc_2023} &
Data in transit, asset protection, governance, operational security, personnel,
supply chain, administration, audit &
Explicitly links cloud suitability to evidence, governance, and administrative
control rather than geography alone. \\[4pt]

Dubai (UAE) &
DESC CSP Security Standard~\cite{desc_2026} &
Mandatory requirements for CSPs serving Dubai government and semi-government
entities &
Shows a jurisdictional model where government cloud supply is conditioned on a
local certification and surveillance regime. \\[4pt]

Saudi Arabia &
NCA Cloud Cybersecurity Controls
(CCC--2:2024)~\cite{nca_2025} &
Minimum cloud cybersecurity requirements for providers and tenants, updated for
data-localisation requirements &
Demonstrates how sovereignty concerns are being translated into national
cloud-control obligations, including localisation-sensitive updates. \\[4pt]

Canada &
GC Digital Sovereignty Framework; GC Cloud Guardrails; data sovereignty white
paper~\cite{gc_digital_sov_2025,gc_guardrails_2024,gc_data_sov_2018} &
Operational resilience, institutional control, legal and supply-chain risk,
mandatory baseline controls &
Directly supports the move from location-centric sovereignty to control,
assurance, and resilience across government operations. \\[4pt]

Singapore &MTCS SS~584:2020 and TR~62 cloud outage incident
response~\cite{imda_2026} &
Tiered cloud assurance and explicit outage-response guidance for cloud
environments &
Combines certification with outage-preparedness, reinforcing that sovereignty
and resilience must be treated together. \\
\bottomrule
\end{tabular}
\end{table*}

Several observations follow from this landscape.

First, mature sovereignty-oriented frameworks rarely stop at residency. Canada explicitly extends the concept into operational resilience and institutional control~\cite{gc_digital_sov_2025}. France binds trusted cloud qualification to technical, operational, and legal conditions~\cite{anssi_cloud_2026}. The UK NCSC principles ask evaluators to inspect governance, administration, supply chain, and audit evidence~\cite{ncsc_2023}. Dubai and Saudi Arabia both translate national cyber policy into concrete CSP obligations~\cite{desc_2026,nca_2025}.

Second, the European landscape is bifurcated between formal certification and ecosystem trust. The EU Cybersecurity Certification Framework provides EU-wide scheme machinery and assurance levels, while the EUCS cloud scheme remains a candidate scheme under development~\cite{eu_ccf_2026,enisa_eucs_2020}. In parallel, Gaia-X has evolved a trust and labelling framework aimed at interoperable, federated, and policy-aware ecosystems, including criteria linked to transparency, portability, and ``European Control''~\cite{gaiax_home_2026,gaiax_fw_2026}. This combination matters because many sovereign services will operate through federated ecosystems rather than purely national stacks.

Third, Europe is now using procurement as a sovereignty lever rather than relying only on certification discourse. The Commission's Cloud Sovereignty Framework defines eight sovereignty objectives spanning strategic, legal and jurisdictional, data and AI, operational, supply chain, technology, security and compliance, and environmental dimensions~\cite{ec_csf_2025}. It then applies minimum SEAL levels and a separate sovereignty score that contributes to tender quality evaluation~\cite{ec_press_2025,ec_csf_2025}. This is one of the clearest current examples of Sovereign~2.0 being quantified, audited, and commercialised as an award criterion.

Fourth, continuity and outage handling are becoming sovereignty concerns in their own right. Singapore's MTCS is a cloud-security standard, but IMDA’s associated technical reference on cloud outage incident response recognises that cloud assurance must be paired with disciplined outage response and disaster recovery expectations~\cite{imda_2026}. The 2026 Middle East incident strongly reinforces that point.

Despite this progress, most existing frameworks remain incomplete when evaluated against crisis conditions. Certification schemes and policy baselines tend to emphasise steady-state compliance, but provide limited guidance on sovereign authority during cross-border disruption, provider unavailability, or emergency migration. In particular, few frameworks explicitly define who retains decision rights over failover, privileged access, or recovery execution when normal jurisdictional and operational assumptions no longer hold.

This gap reinforces the need for a control-plane-centric model of sovereignty that treats governance, operational authority, and evidence generation as first-class design requirements. Without explicit control over these elements, organisations cannot reliably execute recovery or maintain sovereign authority under disruption conditions.

\begin{table*}[t]
\centering
\caption{Comparison of Sovereignty Capabilities Across Models}
\label{tab:sovereignty_comparison}
\begin{tabular}{p{3.5cm} p{3.2cm} p{4cm} p{4cm}}
\hline
\textbf{Capability} & \textbf{Sovereign 1.0 (Residency / Localisation)} & \textbf{Existing Frameworks (e.g., SecNumCloud, EUCS, NCSC, C5)} & \textbf{Sovereign 2.0 (This Paper)} \\
\hline

Data location control & Strong emphasis on in-country storage and processing & Explicitly defined and auditable residency and jurisdictional constraints & Preserved, but treated as necessary rather than sufficient \\

Control-plane authority (governance, admin, policy) & Limited; often implicit or delegated to provider control planes & Partially addressed through governance and administrative requirements & Explicit requirement that governance and privileged control remain under sovereign authority \\

Crisis-time decision authority (failover, migration, recovery) & Not addressed; assumes steady-state conditions & Limited guidance; typically unspecified under cross-border disruption & Explicit requirement for sovereign control over recovery, failover, and emergency actions \\

Operational independence under disruption & Not considered; assumes provider availability & Addressed indirectly via resilience and continuity controls & Explicit requirement to operate and recover services without non-sovereign intervention \\

Evidentiary control (logs, audit, traceability) & Minimal focus beyond compliance artefacts & Strong emphasis on auditability and logging in steady state & Requirement for sovereign access to tamper-evident evidence under both steady-state and crisis conditions \\

Supplier and support-path sovereignty & Not addressed & Partially addressed through supply-chain and personnel controls & Explicit requirement to govern and constrain external support pathways and dependencies \\

Service-boundary coverage (SaaS, identity, telemetry, AI) & Focused on infrastructure and data plane & Expanding coverage, but often scoped to primary cloud services & Explicit inclusion of full service boundary, including external dependencies and control-plane components \\

Cryptographic trust and key control & Implicit or provider-managed & Addressed through key management and encryption requirements & Explicit requirement for sovereign key custody, trust-anchor control, and crypto transition governance \\

Post-quantum readiness & Not addressed & Emerging consideration, not yet systematically integrated & Treated as a foundational sovereignty control for long-lived services \\

Sovereignty under disruption (combined stress conditions) & Not addressed & Limited treatment; primarily steady-state assurance & Central design objective: maintaining control, evidence, and trust under disruption \\

\hline
\end{tabular}
\end{table*}

Sovereignty in this paper is evaluated in terms of control retention: the extent to which governance authority, operational capability, evidence, and trust remain enforceable within the sovereign boundary under both steady-state and disruption conditions.

Existing sovereignty frameworks have significantly advanced beyond pure residency requirements, particularly in areas such as auditability, governance, and supply-chain control. However, when evaluated against the threat model defined in Section II, these frameworks remain primarily optimised for steady-state compliance rather than sovereign control under disruption. Table~\ref{tab:sovereignty_comparison} summarises this distinction.

\section{A Three-Layer Risk-Assurance Model}
\label{sec:assurance}

The principal weakness of many sovereignty programmes is that they stop at policy intent rather than demonstrable control outcomes. Risk registers and procurement clauses state what \emph{should} happen, but they do not reliably prove what \emph{is} happening under steady-state or crisis conditions. To close that gap, this paper proposes a three-layer risk-assurance model: governance assurance, operational assurance, and technical assurance, illustrated in Fig.~\ref{fig:model}.

\begin{figure}[!t]
\centering
\fbox{%
\begin{minipage}{0.9\columnwidth}
\small
\textbf{Drivers of urgency}\\
Kinetic disruption of cloud regions; extraterritorial legal exposure;
service-boundary sprawl through APIs, SaaS, telemetry, identity federation,
AI services, and remote support dependencies.

\medskip\hrule\medskip

\textbf{Layer 1: Governance assurance}\\
Sovereignty profiles by data and service class; jurisdictional acceptability;
decision rights; exception handling; supplier and contract requirements;
continuity thresholds.

\medskip\hrule\medskip

\textbf{Layer 2: Operational assurance}\\
Privileged access governance; incident command; disaster recovery rehearsals;
release and configuration control; evidence production; emergency operating
authority.

\medskip\hrule\medskip

\textbf{Layer 3: Technical assurance}\\
Identity boundaries; encryption and key control; egress and tokenisation; sovereign telemetry; continuous control monitoring; crypto agility and protocol transition readiness.

\medskip\hrule\medskip

\textbf{Management sovereignty control domains}\\
Governance and decision rights $\cdot$ Identity and PAM $\cdot$ Keys and trust anchors $\cdot$ Data lifecycle and egress $\cdot$ Observability and evidence $\cdot$ Supplier and support control

\medskip\hrule\medskip

\textbf{Sovereign service delivery model}\\
Jurisdiction-specific teams $\cdot$ Segregated DevSecOps pipelines $\cdot$ Controlled vendor support paths $\cdot$ Federated continuity runbooks $\cdot$ Local incident and crisis authority $\cdot$ Evidence-as-a-service for auditors and regulators

\medskip\hrule\medskip

\textbf{Foundational trust layer: post-quantum-ready TLS and PKI}\\
Crypto inventory $\cdot$ Hybrid TLS~1.3 transition $\cdot$ Sovereign key custody $\cdot$ Certificate and trust-anchor governance $\cdot$ Protection against harvest-now-decrypt-later exposure.
\end{minipage}}
\caption{Management sovereignty as a three-layer assurance problem delivered through sovereign operating controls and anchored by post-quantum-ready trust.}
\label{fig:model}
\end{figure}

\subsection{Governance Assurance}

Governance assurance defines the rules of the game. It answers what must be
sovereign, why it must be sovereign, who decides, and how exceptions are
governed. Relevant artefacts include data-classification rules, sovereignty
profiles by service class, supplier policies, foreign-law exposure assessments,
exception approvals, and continuity thresholds.

This layer should establish at least four outcomes. First, the organisation
should be able to identify which data classes, keys, identity functions, logs,
and operational roles require sovereign treatment. Second, it should know which
jurisdictions, providers, and support pathways are acceptable for each service
class. Third, it should define emergency decision rights so that crisis actions
are pre-authorised rather than improvised. Fourth, it should create contractual
and policy hooks that obligate external providers to support evidence generation,
incident notification, lawful-access transparency where permitted, and
service-continuity cooperation.

\subsection{Operational Assurance}

Operational assurance is where sovereignty either succeeds or fails. It covers
how systems are actually run: privileged access workflows, just-in-time
elevation, segregation of duties, software release approvals, configuration
baselines, incident command, crisis communications, break-glass procedures,
backup restoration, and cross-region failover exercises.

The 2026 AWS Middle East disruptions make the need for operational assurance
concrete. A sovereign claim has limited value if an organisation cannot execute
an approved migration, recover from remote backups, reroute traffic, or maintain
locally governed incident command under pressure. Operational assurance therefore
requires evidence such as access logs, approval records, rehearsal results,
mean-time-to-recover metrics, alternate-region readiness, and supplier
participation in continuity tests.

\subsection{Technical Assurance}

Technical assurance provides the mechanisms that make governance and operations
enforceable. It includes sovereign identity boundaries, transport and storage
encryption, key management, tokenisation, egress controls, observability
pipelines, policy enforcement points, continuous control monitoring,
tamper-evident logging, and crypto agility.

Technical assurance is also where post-quantum migration belongs. Sovereign control is partly a matter of how trust is established and rotated. If an organisation cannot inventory where vulnerable public-key cryptography is used or update its TLS and PKI dependencies, or cannot ensure sovereign governance of keys and certificate authorities, then its trust boundary remains externally fragile.  Together, these three layers provide a structured mechanism for evaluating and enforcing management sovereignty. They translate abstract sovereignty requirements into observable and testable control outcomes across governance, operations, and technical implementation.

\section{Sovereign Service Delivery in Federated Environments}
\label{sec:federated}

A realistic sovereignty model must assume that many organisations will continue to depend on global providers, shared infrastructure, and specialised suppliers. For that reason, management sovereignty should not be interpreted as a requirement to build every layer domestically. Rather, it should be treated as an operating model that distinguishes which functions may be federated and which must remain sovereign.

In practice, sovereign service delivery requires changes to core operating assumptions.

First, privileged access cannot be globally ambient. Administrative access should be jurisdiction-scoped, time-bounded, approved, monitored, and forensically attributable. Second, support pathways must be treated as control boundaries. The ability of an external provider to ``help'' with a service is itself a sovereignty decision that should be policy-driven and auditable. Third, DevSecOps pipelines, configuration repositories, and production-signing processes should be segmented in proportion to data class and continuity criticality. Fourth, continuity plans should assume not only conventional outages, but also loss of region, loss of provider pathway, or temporary suspension of normal cross-border operations. Fifth, evidence collection should be designed as an operational product, not an afterthought.

This federated approach aligns with several of the frameworks surveyed above. Canada's digital sovereignty framework emphasises legal, supply, and technical controls alongside continuity and resilience~\cite{gc_digital_sov_2025}. The NCSC principles place strong emphasis on administrative security, personnel trustworthiness, supply-chain security, and customer audit information~\cite{ncsc_2023}. Singapore's outage-response guidance reinforces that cloud assurance is incomplete without disciplined incident and continuity practice~\cite{imda_2026}.

\subsection{Example Implementation Pattern}

To make management sovereignty concrete, a representative sovereign service architecture should implement control-plane isolation and evidence generation across key domains.

Identity and privileged access should be enforced through jurisdiction-scoped privileged access management (PAM), with just-in-time elevation, multi-party approval, and full session recording. Administrative actions must be attributable and governed within the sovereign boundary.

Cryptographic trust should be anchored in sovereign key custody, with hardware security modules (HSMs) under local control, sovereign certificate authorities, and policy-governed key lifecycle management. External trust dependencies should be minimised or explicitly governed.

Observability pipelines should retain primary logs, traces, and alerts within the sovereign domain, with export controls applied to downstream analytics or external monitoring services. Logs should be tamper-evident and accessible for regulatory inspection.

Data egress should be mediated through policy enforcement points, including API gateways, tokenisation, and allowlisting, ensuring that both bulk data and operational metadata flows are governed.

Continuity architecture should include pre-authorised failover patterns, jurisdiction-aware disaster recovery runbooks, and clearly defined incident command authority, enabling rapid migration or recovery without requiring external approval under crisis conditions.

Together, these patterns illustrate how management sovereignty can be implemented as an enforceable control system rather than a declarative policy posture.

\section{External Services as Sovereignty Leakage Paths}
\label{sec:leakage}

In many environments, the largest sovereignty exposure is not the primary compute platform but the surrounding services. SaaS collaboration suites, observability providers, security tooling, code-hosting platforms, AI model endpoints, ticketing systems, and identity brokers all process operationally significant data. Some services process content, while others handle metadata and administrative context. 

Management sovereignty treats these services as managed dependencies rather than peripheral tools. Four design principles follow.

The first is \emph{classification-driven service use}. Not all service classes should be allowed to handle all data classes, and the rules should extend to telemetry, prompts, model outputs, and administrative metadata. The second is \emph{egress discipline}. API gateways, data loss prevention, tokenisation, and allowlisting must be applied not only to bulk transfers but also to streaming and support paths. The third is \emph{evidence discipline}. External services should supply immutable audit evidence, retention controls, and notification pathways. The fourth is \emph{recoverability discipline}. Organisations should know how to continue operating when a peripheral dependency becomes unavailable, legally inaccessible, or politically unacceptable.

This is a primary source of sovereignty leakage in otherwise well-architected platforms. A sovereign hosting enclave is weakened if its logs are exported to a non-sovereign platform, if its identity plane is anchored elsewhere, or if emergency support requires foreign privileged access that cannot be constrained or independently evidenced.

\section{Post-Quantum TLS as a Sovereignty Control}
\label{sec:pqtls}

\subsection{Why TLS Belongs Inside the Sovereignty Model}

TLS is commonly treated as transport plumbing. In sovereign service delivery it should be treated as a trust control. APIs, identity federation, administrative consoles, service meshes, data replication links, and inter-agency exchanges all depend on TLS-mediated trust relationships. If those trust relationships cannot be inventoried, updated, and governed under sovereign control, then neither the confidentiality nor the continuity of the service boundary is fully sovereign.

\subsection{Why the Timing Is Immediate}

NIST is explicit that the transition to post-quantum cryptography should begin now. Its plain-language guidance warns that encrypted data is already vulnerable to ``harvest now, decrypt later'' collection strategies for secrets with long confidentiality lifetimes~\cite{nist_pqc_what_2026}. The broader NIST PQC programme notes that the first three standards are ready for implementation now~\cite{nist_pqc_2026}. NIST finalised FIPS~203, FIPS~204, and FIPS~205 in August~2024~\cite{nist_fips_2024,fedregister_fips_2024}. NIST's migration project and draft transition guidance further stress the need to inventory quantum-vulnerable public-key dependencies and build roadmaps across hardware, software, services, and protocols~\cite{nist_nccoe_2026,nist_ir8547_2024}.

This matters directly for sovereign services because identity and inter-service trust often protect data whose value outlives the current cryptographic era: citizen records, health information, justice data, operational telemetry, and critical-infrastructure interactions. Where confidentiality lifetimes are long, waiting for a complete post-quantum market transition is itself a sovereignty risk.

\subsection{A Pragmatic Transition Pattern}

The IETF's work on hybrid key exchange in TLS~1.3 provides a practical transition pattern: combine classical and post-quantum key establishment so that the session remains secure unless all component mechanisms are broken~\cite{ietf_hybrid_2025}. This approach is attractive for sovereign service delivery because it supports staged migration without requiring an instantaneous replacement of the installed base.

A sovereignty-oriented TLS transition should therefore include at least four workstreams. The first is a \emph{cryptographic inventory}: every externally reachable service, internal API, VPN, load balancer, service mesh, HSM, certificate authority, and identity component that depends on public-key cryptography should be identified. The second is \emph{crypto agility}: systems should be able to change key-establishment and signature algorithms without wholesale redesign. The third is \emph{hybrid transition}: organisations should test and phase in hybrid TLS where interoperability and platform support make it feasible. The fourth is \emph{sovereign key custody}: certificate issuance, revocation, rotation, HSM policy, root and intermediate trust anchors, and approval workflows should remain under sovereign governance.

In this sense, post-quantum TLS is not a cryptographic side issue. It is part of the same control problem as privileged access, logging, and continuity. It determines whether the service boundary can retain trustworthy communications and identity under future cryptanalytic conditions.

\section{Discussion}
\label{sec:discussion}

Three broader implications follow from the argument developed above.

First, sovereign service delivery should be framed as a spectrum of control outcomes rather than an all-or-nothing infrastructure attribute. Full autonomy is rarely realistic, and the Government of Canada explicitly recognises that complete digital autonomy is impossible in a globally connected environment~\cite{gc_digital_sov_2025}. A more useful design question is: which functions must remain sovereign, at which assurance level, under which disruption assumptions?

Second, the most credible sovereignty models are those that combine legal, operational, and technical controls. Residency-only models fail under foreign legal compulsion. Operational-only models fail when evidence is weak or cryptographic trust is externally anchored. Technical-only models fail when supplier contracts, emergency decision rights, or continuity authority are vague. Management sovereignty is valuable precisely because it compels these dimensions to be designed together.

Third, the path forward is likely to be federated rather than isolationist. That conclusion is consistent with recent descriptions of ``Sovereignty~2.0,'' which emphasise strategic autonomy through selective interdependence rather than complete autarky~\cite{wef_2025}. The standards landscape already points in this direction. EU certification and Gaia-X emphasise common assurance and federation~\cite{eu_ccf_2026,gaiax_fw_2026}. The European Commission's Cloud Sovereignty Framework goes further by embedding sovereignty metrics directly into procurement~\cite{ec_press_2025,ec_csf_2025}. Canada stresses interoperability with trusted partners alongside control and resilience~\cite{gc_digital_sov_2025}. Singapore couples tiered assurance with outage response~\cite{imda_2026}. Even highly trust-sensitive frameworks such as SecNumCloud address how commercial cloud can be made trustworthy under stringent conditions rather than assuming that only wholly national stacks are acceptable~\cite{anssi_cloud_2026}.

This paper is a design and governance analysis rather than an empirical evaluation. The AWS Middle East disruptions are used as a contemporary forcing function rather than as a comprehensive case study. Future work should apply the proposed model to real-world deployments to validate its effectiveness across different sectors and threat conditions. In addition, some key public schemes are still evolving; most notably, the EUCS cloud scheme remains a candidate scheme rather than a finalised European sovereign-cloud certification~\cite{enisa_eucs_2020}. These limitations do not undercut the central argument, but they do mean that management sovereignty should be refined further through operational field studies, assurance metrics, and sector-specific deployment patterns.

Management sovereignty is not without trade-offs. Implementing jurisdiction-scoped control-planes, sovereign key custody, and independent observability increases architectural complexity and operational overhead. It may also constrain the use of globally integrated SaaS platforms and limit certain economies of scale offered by hyperscale providers.

In addition, enforcing sovereign control over support pathways and administrative access can introduce latency in incident response if not carefully designed. These trade-offs require deliberate balancing between sovereignty, cost, performance, and operational agility.

For this reason, management sovereignty should be applied proportionally, with higher assurance levels reserved for services with greater sensitivity, regulatory exposure, or continuity criticality.
\section{Conclusion}
\label{sec:conclusion}

The sovereign-cloud debate has entered a new and explicitly operational phase. The combination of kinetic stress on cloud infrastructure, ongoing extraterritorial legal exposure, expanding service boundaries, and the post-quantum transition means that sovereignty can no longer be reduced to data residency or local hosting claims. The 2026 AWS Middle East disruptions demonstrated that regional cloud dependency is a continuity problem as much as a compliance problem~\cite{aws_apr5_2026,aws_rss_2026,ap_chan_2026}. Contemporary government and standards frameworks increasingly reflect the same shift, whether through operational resilience and institutional control in Canada, trusted-cloud qualification in France, governance and audit expectations in the UK, or cloud control baselines in Dubai, Saudi Arabia, and Singapore~\cite{anssi_cloud_2026,ncsc_2023,desc_2026,nca_2025,gc_digital_sov_2025, imda_2026}.

This paper has argued for management sovereignty as the appropriate next-step concept. In the terminology proposed here, Sovereign~1.0 remains necessary but insufficient: location still matters, but it does not settle the questions of authority, evidence, recovery, and trust. Sovereign~2.0 adds those dimensions by evaluating whether governance, privileged access, cryptographic trust, data lifecycle and egress, observability, supplier dependency, and recovery authority remain under sovereign control and can be continuously evidenced. Under that model, risk assurance is not a supporting activity but the mechanism by which sovereign claims become credible. The European Commission's Cloud Sovereignty Framework is important precisely because it shows how a jurisdiction can convert those claims into minimum assurance thresholds and procurement scoring rather than leaving them at the level of policy rhetoric~\cite{ec_press_2025,ec_csf_2025}. Post-quantum-ready TLS and PKI are likewise not future nice-to-haves, but foundational trust controls that help keep sovereign service boundaries viable over the confidentiality lifetime of the data they protect.

The practical implication is clear: sovereign services must be engineered and operated as evidence-backed control systems capable of maintaining authority, trust, and continuity under both jurisdictional pressure and physical disruption.

\section{Future Work and Research Agenda}

The model of Sovereign 2.0 and management sovereignty defined in this paper establishes a foundation for a broader research and implementation programme. Several directions follow directly from this work.

First, there is a need for formalised measurement and assessment. Future work should develop sovereignty metrics and maturity models that quantify control retention across governance, operational, technical, and cryptographic domains. Such metrics would enable organisations and regulators to evaluate the degree to which sovereign authority is maintained under both steady-state and disruption conditions, and to compare alternative architectures and provider models.

Second, the translation of management sovereignty into enforceable procurement and assurance mechanisms remains an open problem. While emerging frameworks introduce sovereignty objectives and scoring, further work is required to define verifiable control requirements, audit artefacts, and continuous assurance mechanisms that can be embedded in contracts, certification schemes, and regulatory oversight.

Third, reference architectures and implementation patterns should be developed and validated across representative environments. This includes formalising sovereign control-plane isolation patterns, evidence pipelines, jurisdiction-scoped identity and access models, and continuity architectures capable of operating under region loss, supplier unavailability, or cross-border constraint.

Fourth, sector-specific applications of management sovereignty require further exploration. Critical infrastructure, public-sector systems, and regulated industries exhibit different risk profiles, legal exposures, and continuity requirements. Applying the model across these domains will help refine proportionality, assurance levels, and control prioritisation.

Finally, the role of cryptographic sovereignty in long-lived systems warrants deeper investigation. The transition to post-quantum cryptography introduces new requirements for crypto agility, key governance, and trust-anchor control. Future work should examine how sovereign control over cryptographic mechanisms can be maintained across heterogeneous platforms and evolving standards over multi-decade service lifecycles.

Together, these directions position Sovereign 2.0 not as a static framework, but as the basis for an evolving discipline of sovereign service delivery, spanning architecture, policy, assurance, and operational practice.

\balance
\bibliographystyle{IEEEtran}
\bibliography{references}

\end{document}